\documentclass[12pt,a4paper]{article}
\usepackage[symbol]{footmisc}

\usepackage{parskip}
\usepackage[margin=0.7in]{geometry}

\usepackage{amsmath}
\usepackage{physics}
\usepackage{kmath}
\usepackage{kerkis} 

\usepackage[english]{babel}
\usepackage{amstext}
\usepackage{amssymb}
\usepackage{graphicx}
\usepackage{dsfont}
\usepackage{float}
\usepackage{xcolor}
\usepackage{textcomp}
\usepackage{listings}
\usepackage{relsize}
\usepackage{multicol}
\usepackage{subcaption}
\usepackage{hyperref}
\hypersetup{
	colorlinks=true,
	linkcolor=black,
	filecolor=black,      
	urlcolor=black,
	citecolor=blue,
}
\usepackage{physics}
\usepackage{url}

\usepackage{algorithm}
\usepackage{algpseudocode}
\usepackage{pifont}

\usepackage{bbm}
\usepackage{mathtools}
\usepackage{hhline}
\usepackage{tabularx}

\usepackage{verbatim}

\usepackage{footnote}
\usepackage{footmisc}

\begin{document}

\begin{center}

{{{\Large\bf Estimation of the effective reproduction number for SARS-CoV-2 infection during the first epidemic wave in the metropolitan area of Athens, Greece }}}\\

\vspace{0.5cm}
Konstantinos Kaloudis\textsuperscript{1}, 
George A. Kevrekidis\textsuperscript{2},
Helena C. Maltezou\textsuperscript{3},
Cleo Anastassopoulou\textsuperscript{4}, 
Athanasios Tsakris\textsuperscript{4}, 
Lucia Russo\textsuperscript{1,}\footnote[1]{Corresponding author, E-mail address: l.russo@irc.cnr.it}

\vspace{0.2cm}
\end{center}
\centerline{\footnotesize \phantom{0}\textsuperscript{1} National Research Council, Science and Technology for Energy and Sustainable Mobility, Naples, Italy}
\centerline{\footnotesize \phantom{0}\textsuperscript{2} Department of Mathematics and Statistics, University of Massachusetts Amherst, MA, USA} 
\centerline{\footnotesize \phantom{0}\textsuperscript{3}
Directorate of Research, Studies and Documentation, National Public Health Organization, Athens, Greece}            
\centerline{\footnotesize \phantom{0}\textsuperscript{4} Department of Microbiology, Medical School, University of Athens, Athens, Greece}
\vspace{0.2in}
\begin{abstract}
Herein, we provide estimations for the effective reproduction number $R_e$ for the greater metropolitan area of Athens, Greece during the first wave of the pandemic (February 26-May 15, 2020). For our calculations, we implemented, in a comparative approach, the two most widely used methods for the estimation of $R_e$, that by Wallinga and Teunis and by Cori et al. Data were retrieved from the national database of SARS-CoV-2 infections in Greece. Our analysis revealed that the expected value of $R_e$ dropped below 1 around March 15, shortly after the suspension of the operation of educational institutions of all levels nationwide on March 10, and the closing of all retail activities (cafes, bars, museums, shopping centres, sports facilities and restaurants) on March 13. On May 4, the date on which the gradual relaxation of the strict lockdown commenced, the expected value of $R_e$ was slightly below 1, however with relatively high levels of uncertainty due to the limited number of notified cases during this period. Finally, we discuss the limitations and pitfalls of the methods utilized for the estimation of the $R_e$, highlighting that the results of such analyses should be considered only as indicative by policy makers. 
\end{abstract}

\section{Introduction}
To mitigate the devastating effects of the coronavirus disease 2019 (COVID-19) pandemic that was spreading rapidly across Europe in the first months of 2020, countries imposed the only available, centuries-old public health interventions for contagious infectious diseases of major public health importance: shutdowns and social-distancing. The rapid spread and intensity of the pandemic in neighboring Italy, prompted immediate and drastic measures in Greece through fears that the national health system would collapse. The early taken  mitigation measures were proven effective at those early stages, resulting in low numbers of hospitalizations and deaths, both in relative and absolute measure. The situation changed in the autumn of 2020, following the lift of containment measures and the opening of the country to tourists during the summer. The second COVID-19 epidemic wave that started in September 2020 and is currently reaching its peak, hit Greece hard: cases rose sharply, hospitalizations strained the health system, and the death ratio spiked (indicatively for the period Nov 1- Dec 5: Greece: 30 deaths per 1,000 confirmed cases compared to 38 in Belgium, 20.7 in the United Kingdom, 20 in France, 20 in Italy, 18.5 in Spain, 13.7 in Portugal  and 13.2 in Austria)  \cite{WHO2020dash}.\par  
The scientific response to the pandemic has been intense and multidisciplinary in an effort to understand and access a wide range of attributes, extending from the biological characteristics of the virus to the effects of social-distancing measures and human behavior. However, since COVID-19 is a new disease caused by a novel viral pathogen, several unknowns and uncertainties remain to be elucidated in order to fully comprehend its complexities and end the pandemic. Mathematical modelling provides indispensable tools in this respect. One such tool, or compass with which to navigate the unknown waters of the pandemic as safely as possible, is the effective reproduction number $R_e$, i.e. the ratio of secondary cases from a contagious person 
 \cite{gunther2020nowcasting, Nishiura_2009}. The real-time estimation of $R_e$ has been key for the assessment of the evolution of the pandemic and the application of public health policy measures 
 \cite{gunther2020nowcasting, Inglesby_2020}.  Its estimation relies on both data availability and such epidemiological features as the incubation period, serial interval, generation time, delays between infection and case observation and, importantly, the interlinked scales under-reporting and test policy. However, there is not a one-fits-all method for its estimation \cite{gostic2020practical}. Two main approaches are followed for the estimation of $R_e$: (a) the model-based, where one extracts $R_e$ by first fitting the parameters of compartmental dynamic models \cite{wallinga2007generation,bettencourt2008real,Diekmann_2009, Arias2009,Ridenhour_2014,Russo_2020}, and (b) the time-series-based analysis  \cite{wallinga2004different,cori2013new,gostic2020practical}. For a review of the above methods and suggestions about their applicability see \cite{gostic2020practical}).\par
In this work, we used the two most widely used approaches, the one developed by Wallinga and Teunis \cite{wallinga2004different} and the one by Cori et. al. \cite{cori2013new} to estimate the $R_e$ during the period February 26-May 15 2020 in Attica (the metropolitan area of Athens and its suburbs). Lytras et al. \cite{lytras2020improved} have reported an estimation of the $R_e$ for the whole country by extending the Cori et al. method using a Bayesian approach and investigated the evolution of $R_e$ estimates with respect to the mitigation measures. Here, we focus in Attica, thus relaxing the assumption of the heterogeneity due to both geography but more importantly the mobility between regions, which comprises a few urban, and many quite diverse rural areas, ranging from remote mountain villages to small seaside towns on the islands. This heterogeneity sets additional uncertainty in the estimations of time series analyses, especially during the unusual circumstances of a lockdown. Thus, our study attempts to provide an estimation of the $R_e$ and capture the dynamics of the evolving epidemic during the period of the first lockdown in the specific metropolitan area of Athens (the region of Attica), where approximately half the population of Greece is concentrated. 

\section{The Epidemiological Data from Attica, Greece}

\subsection{Surveillance of SARS-CoV-2 infection}  
SARS-CoV-2 infection is notifiable in Greece. Case-based data are collected from laboratories diagnosing SARS-CoV-2 infection by real-time reverse-transcription polymerase chain reaction (RT-PCR). In addition, physicians notify laboratory-confirmed SARS-CoV-2 cases via the mandatory notification form. A passive comprehensive system for hospitalized cases is also in place, collecting data on a daily basis about admission to intensive care unit (ICU), incubation, complications, and outcome. Active, massive testing, regardless of symptoms, is also performed for containment purposes and in the context of investigation of clusters in closed settings or specific populations (long-term care facilities, Roma populations, refugee hosting centers, repatriates). 
\subsection{Contact tracing and measures implemented}
Active and exhaustive contact tracing was implemented upon diagnosis of the first case throughout the first epidemic wave in Greece. Close contacts of SARS-CoV-2- infected cases were instructed to stay isolated for 14 days after their last contact. In case of onset of symptoms, the contacts were advised to attend a healthcare facility for testing.  
\subsection{Data collection} 
We retrieved data on SARS-CoV-2 infected cases who resided permanently in Attica of a total population of 4m people.  Data were retrieved from the national database of SARS-CoV-2 infected cases. The study period extended from February 26, 2020 (diagnosis of the first COVID-19 case in Greece) through May 15, 2020. Data on clinical course and outcome of patients were updated on June 5, 2020. The total number of cases was 1,645, out of which 268 were asymptomatic. 
\subsection{Definitions} 
SARS-CoV-2 infection was defined as a laboratory-confirmed infection with SARS-CoV-2 regardless of symptoms. COVID-19 was defined as a case with signs and symptoms compatible with COVID-19 and laboratory-confirmed SARS-CoV-2 infection. Laboratory-confirmed SARS-CoV-2 infection was defined as a case tested positive by RT-PCR. COVID-19 associated death was defined as death of a COVID-19 case with no period of complete recovery between the illness and death and in the absence of a clear alternative cause of death.

\section{Methodology}

Our aim was to provide estimations of the effective reproduction number $R_e(t)$ during the period February 26 - May 15, for which detailed data for the greater metropolitan area of Athens were available from the national surveillance database of SARS-CoV-2 infections (National Public Health Organization, Athens). This period contains the lockdown period of March 23 $-$ May 4.

It is well known that $R_e$ is determined by the serial-interval distribution, defined as the time period between manifestation of symptoms in the primary case to time of symptom manifestation in the secondary case \cite{svensson2007note}.\par 
Here, as a first step, we performed an imputation of the data (regarding the date of symptoms onset). Then, we used the two most common methods for the estimation of $R_e$, namely the Wallinga and Tunis \cite{wallinga2004different} and the Cori et al. \cite{cori2013new} method. Finally, we performed a sensitivity analysis with respect to the imputation of the data.\par

\subsection{Methods for estimating $R_e$}
\subsubsection{Approximation of the generation time distribution
}
For the estimation of the effective reproduction number it is necessary to know the distribution of generation times defined as the time-lag between an infection in a primary case and an infection in a secondary case. However this information is generally unknown. Thus, this distribution is usually approximated with the serial interval distribution containing the time-lag between onset of symptoms in a primary case and an infection in a secondary case. For a discussion regarding the differences between the generation time and the serial interval we refer to Svensson (2017) \cite{svensson2007note}.\par
\begin{figure}[H]
	\centering
	\includegraphics[keepaspectratio,scale=0.65]{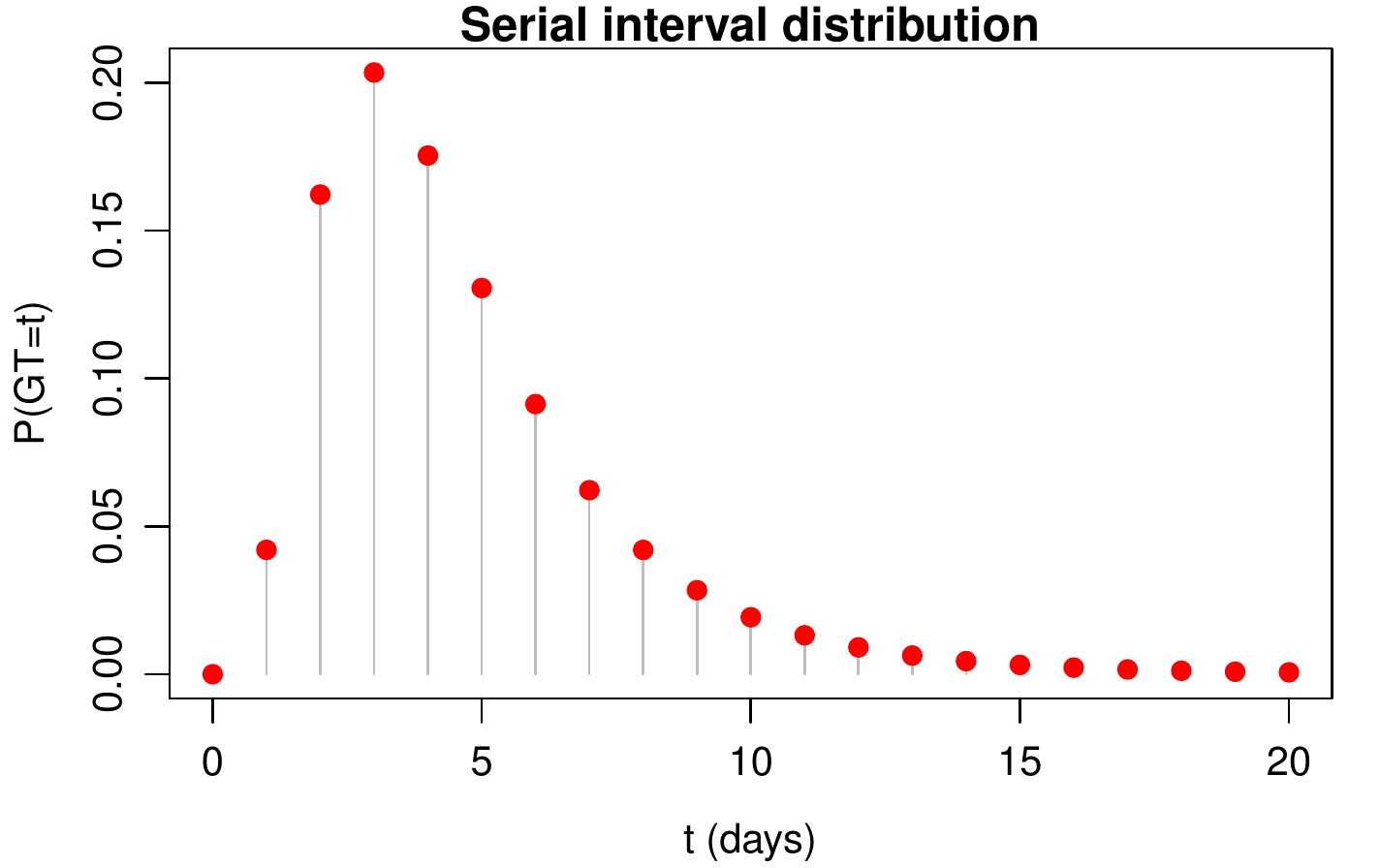}
	\caption{Serial interval distribution, describing the time between symptoms between primary and secondary cases. Its construction was based on a right truncated discretized lognormal distribution with mean and standard deviation equal to 4.7 and 2.9, respectively \cite{nishiura2020serial}. \label{fig4}}
\end{figure}

Here, the reconstruction of the serial interval distribution for COVID-19 was based on the procedure described in Nishiura et al. \cite{nishiura2020serial}. In particular, for its derivation, we used a right truncated (at 20 days) discretized lognormal distribution with mean and standard deviation equal to 4.7 and 2.9, respectively. The resulting serial interval distribution is given in 
Fig. \ref{fig4}.  

\subsubsection{The method of Wallinga and Teunis for estimating $R_e$}
The method proposed by Wallinga \& Teunis (WT) \cite{wallinga2004different} estimates $R_e$ by averaging over all transmission networks compatible with observations and provides likelihood-based estimates using only \textit{pairs} of cases and not the whole infection network which is usually unknown. According to the WT method, the ``who infected whom" pair-wise infection network is approximated in a probabilistic manner from the given
dates of symptom onset. In particular, $p_{ij}$ defines the relative likelihood that a case $j$ is infected by a case $i$ given the time-lag between the onset of symptoms of the two cases ($t_i-t_j$) and it is expressed as:
\begin{align}
    p_{ij}=\frac{w(t_i-t_j)}{\sum_{i\neq k}w_s(t_i-t_k)},
\end{align}
where $w_s$ is the distribution of the generation time, which is specific to the disease.
Based on the above, the effective reproduction number at time (day) $t$ is given by:
\begin{align}
    R_e(t)=\frac{1}{n_j}\sum_{j}R_{e,j}
\end{align}

where $R_{e,j}=\sum_{i}p_{ij}$ is the effective reproduction number for each single case $j$  and $n_j$ denotes the number of all cases $j$ who show
the first symptoms of illness on day $t$ \cite{wallinga2004different}.

Here, we note that a key assumption of the WT model is that the infection network of the observed cases can be built by considering just the notified cases. Thus, this imposes an important limitation regarding its applicability in cases where a large proportion of the population is infected and in the case of many asymptomatic cases. We address this particular limitation in the discussion section.\par

\subsubsection{The method of Cori et al. for estimating $R_e$}

The method proposed by Cori et al. (2013) \cite{cori2013new} and implemented in \cite{coriepiestim} is a Bayesian method suitable for real-time estimation of $R_e$. It was addressed to overcome a main drawback of the WT method, when a real time estimation of $R_e$ is required. In particular, in the method of WT the estimation of $R_e$ at a time $t$ may require incidence data from times later than time  $t$ (for a detailed discussion please refer to \cite{cori2013new}). In fact, the method of Cori et al. provides an estimation of the so-called instantaneous reproduction number, defined as the ratio of the expected number of new infections generated
at time $t$, say $\mathbb{E}[I(t)]$, to the total number of infected
individuals at time $t$, i.e.:
\begin{align}
   R_{e}(t)=\frac{\mathbb{E}[I(t)]}{\sum_{s=1}^tI(t-s)w(s) }
\end{align}
$I(t)$ is the number of new infections at time $t$, and is treated as a random variable and $w_s$ represents the infectivity profile that is the distribution of the infectiousness through time after infection. As the infectivity profile is usually not known this is approximated
by the distribution of the serial interval \cite{cori2013new}.\par
Using a Gamma distribution as a prior to a Bayesian inference procedure, we obtain an estimation for the posterior distribution of $R_{e}(t)$. For more details please refer to the supplementary information in \cite{cori2013new}.
As the estimates of $R_{e}(t)$ can significantly vary over time instances \cite{cori2013new}, what is usually done to obtain smoother results is to apply the above calculations not instantaneously, but within a sliding window of size $t-\tau$: $[t-\tau+1, \cdots, t]$ and calculate the average $R_{e,\tau}$ over this window. The selection of the window is a trade-off, as very long windows lead to oversmoothing and very short ones lead to high levels of noise.

\subsection{Imputation of the Epidemiological Data}
The estimation of $R_e$ in both approaches is based on the knowledge of the symptom onset dates. In the dataset there were 268 (out of 1,645 cases) asymptomatic cases with obviously no symptom onset dates. Thus, for the imputation of the unknown ``symptom'' onset dates for this category (which actually mark the onset of disease transmission), we used an approach by fitting a generalized additive model for location, scale and shape (GAMLSS) \cite{stasinopoulos2007generalized}. For our computations, as suggested by \cite{gunther2020nowcasting,haw2020epidemiological}, it is assumed that the distribution of the delay times $t_d$ (delay from onset of symptoms to the reporting date) is a  Weibull distribution of the form:
$$ t_d \sim \mathcal{W}\left(\mu,\sigma\right),$$

with density $f(t_d ;\mu,\sigma)=\sigma\,\mu\,t_d^{\sigma-1}\exp{-\mu\,t_d^{\sigma}}$ and $\mu>0,\,\sigma>0$ the location and scale parameters, respectively.

In the GAMLSS model both parameters of the Weibull distributions were estimated using the same model, reading:
\begin{equation}
\eta_j = \beta_{0,j} + \sum_{i=1}^{6}\beta_{i,j}\mathcal{I}\left(x_{\text{weekday}}=i\right) + f_{1,j}\left(x_{\text{week}}\right), \,\,\, \eta_{j=1,2} \in \left\{\mu,\sigma\right\}
\end{equation}

In the above, we denote with $\beta_{0,j}$ the intercepts for the location and scale, while $\beta_{i,j},\,\, i=1,\ldots,6$ are the model parameters reflecting the (categorical) effects of the days of the week when each case was reported to the health authorities. The effect of the reporting week is modeled by smoothing penalized splines (P-splines), here denoted by $f_{1,j}(\cdot)$ \cite{eilers2015twenty, gunther2020nowcasting}.

\section{Results}

For the numerical simulations we have used Python (package ``scipy.stats'') \cite{2020SciPy-NMeth} and R \cite{ihaka1996r} programming languages. Our code for the imputation step was based on the code provided by the github public repository\footnote[2]{\url{https://github.com/FelixGuenther/nc_covid19_bavaria}} for the manuscript \cite{gunther2020nowcasting}. For the computations regarding the $R_e$ estimation, we used the R packages ``R0'' \cite{obadia2012r0} and ``EpiEstim \cite{coriepiestim}.

\subsection{Descriptive Statistical Analysis of the Epidemiological Data}
\begin{figure}[H]
	\centering
	\includegraphics[keepaspectratio,scale=0.6]{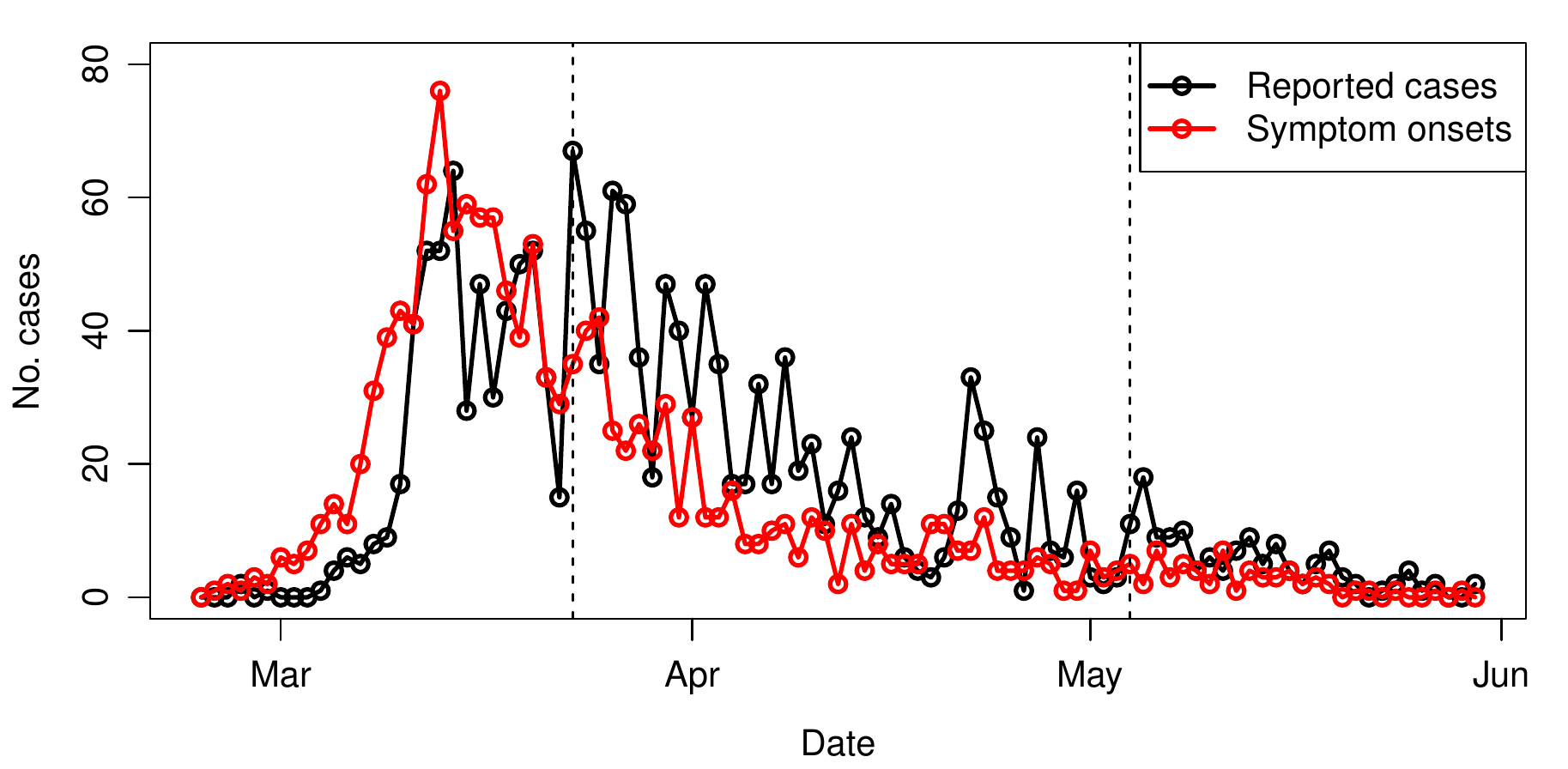}
	\caption{Daily numbers of notified cases and symptom onsets in Attica region, between 25/02/2020 and 30/05/2020. With dashed lines we indicate the begin (23/03) and end (04/05) dates of the general lockdown period.}\label{fig1}
\end{figure}

\begin{figure}[H]
  \begin{subfigure}[b]{0.49\textwidth}
    \includegraphics[width=\textwidth]{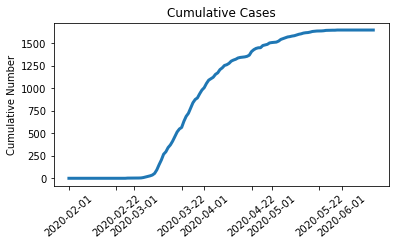}
    \caption{Cumulative SARS-CoV-2 infected cases}
    \label{fig:cumulative_cases}
  \end{subfigure}
  \hfill
  \begin{subfigure}[b]{0.49\textwidth}
    \includegraphics[width=\textwidth]{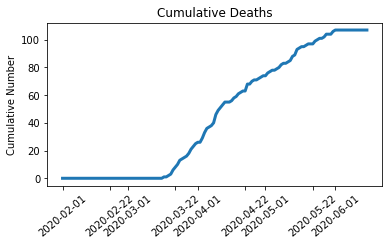}
    \caption{Cumulative Deaths}
    \label{fig:cumulative_deaths}
  \end{subfigure}
  \caption{Cumulative SARS-CoV-2 infected cases (1645 total) and Deaths (107 total) over time.}  \label{fig:cumulative_figures}
\end{figure}

The daily numbers of notified cases and reported symptom onsets are presented in Fig. \ref{fig1}, while  the cumulative numbers of cases and deaths are shown in in Figs. \ref{fig:cumulative_cases} and \ref{fig:cumulative_deaths}, respectively.  

Following the analysis in other studies for the determination of statistical characteristics of COVID-19 (see e.g. \cite{doi:10.1063/5.0013031},\cite{PPR:PPR173058},\cite{Marsland2020.04.21.20073890}, \cite{Nishimoto2020.07.02.20144899}) we have fitted the distributions of time delays between (a) symptom onset and test, (b) symptom onset and death, (c) hospitalization and death, (d) hospitalization and discharge using several types of distributions available in the Python package ``scipy.stats'' \cite{2020SciPy-NMeth} ((a) was directly fitted to a Weibull distribution following the references on onset date imputation in section 3.2). In particular, we used the following distributions\footnote[3]{The associated density functions  can be found at \url{https://docs.scipy.org/doc/scipy/reference/stats.html}.}: Beta, Cauchy, Exponential, EMG, F, Logistic, Gamma, Inverse Gamma, Student's T and non-central Student's T, truncated Exponential and truncated Normal. The parameters of the above distributions were found by minimizing the RSS error between the empirical data and the distribution's pdf. Additionally, there is no evidence that these distributions vary over time throughout our dataset, so we assumed that they are constant during the entire period.\par
We first find the best-fit parameters and then evaluated their moments based on their closed form. We then used bootstrapping to compute their 95\% confidence intervals.

\begin{figure}[h!]
  \begin{subfigure}[b]{0.49\textwidth}
    \includegraphics[width=\textwidth]{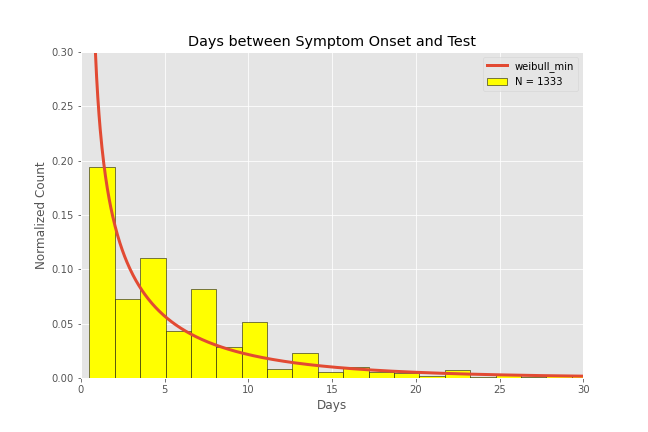}
    \caption{Days between Symptom Onset and Test}
    \label{fig:onset_to_test}
  \end{subfigure}
  \hfill
  \begin{subfigure}[b]{0.49\textwidth}
    \includegraphics[width=\textwidth]{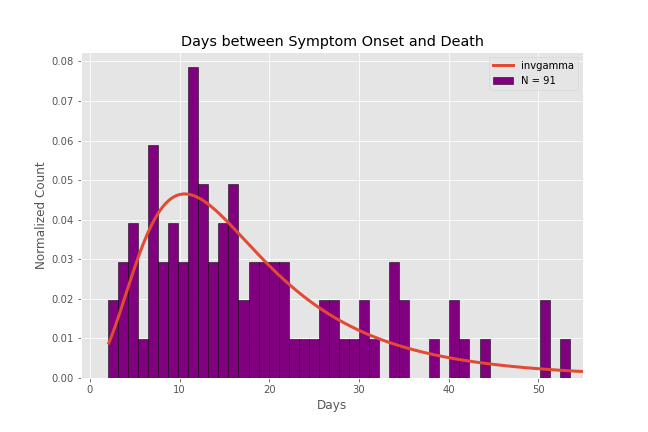}
    \caption{Days between Symptom Onset and Death}
    \label{fig:onset_to_death}
  \end{subfigure}\\
  \begin{subfigure}[b]{0.49\textwidth}
    \includegraphics[width=\textwidth]{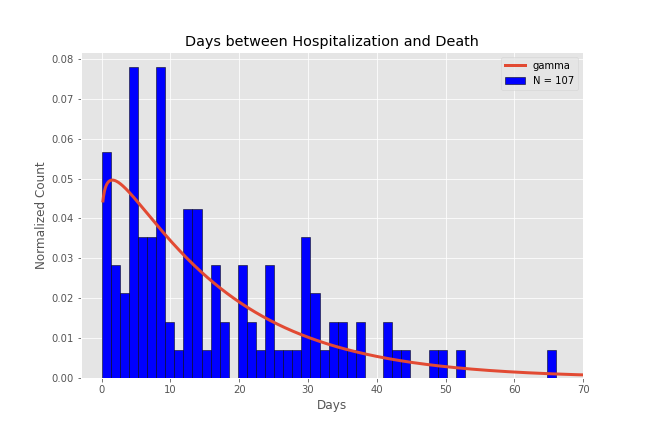}
    \caption{Days between Hospitalization and Death}
    \label{fig:hosp_to_death}
  \end{subfigure}
  \hfill
  \begin{subfigure}[b]{0.49\textwidth}
    \includegraphics[width=\textwidth]{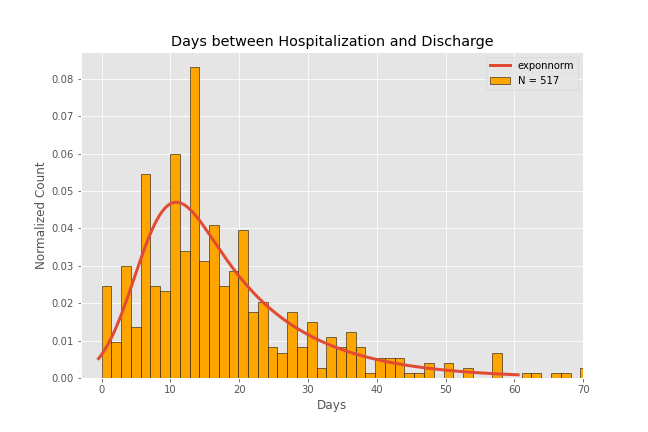}
    \caption{Days between Hospitalization and Discharge}
    \label{fig:hosp_to_discharge}
  \end{subfigure}\\
  \begin{center}
  \begin{subfigure}[b]{0.49\textwidth}
    \includegraphics[width=\textwidth]{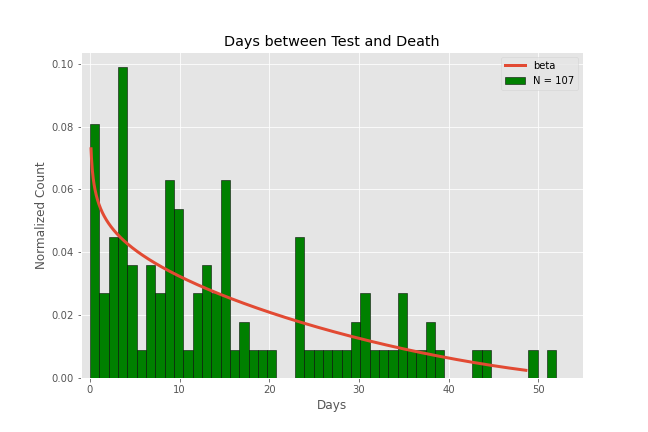}
    \caption{Days between Test and Death}
    \label{fig:test_to_death}
  \end{subfigure}
  \end{center}
  \caption{Histograms of the characteristic time intervals for the Attica dataset and curves of best-fitted distributions. Normalized histograms of characteristic time intervals for the Attica dataset; the red lines depict the best-fit distributions. The specifics about the distribution parameters are summarized in Table \ref{table:dist_details}.}
  \label{fig:distributions}
\end{figure}

In Fig. \ref{fig:distributions} we present the best-fit distribution  and in Table \ref{table:dist_details} we present the fitted parameters along with some relevant statistics.

\begin{table}[h!]
    \begin{tabular}{|p{3cm}||p{2.5cm}|p{2.3cm}|p{2.3cm}|p{2.3cm}|p{2cm}| }
     \hline
     \multicolumn{6}{|c|}{Summary of Fitted Distributions in Figure \ref{fig:distributions}}\\
     \hline
     Data & Distribution&Mean&St. Dev. & Parameters& RSS Error\\
     \hhline{|=|=|=|=|=|=|} 
     Onset to Test (\ref{fig:onset_to_test})&Weibull& 5.6&7.4&$c=0.70$&0.015\\
     \hline
     C.I. (95\%) & &(3.8, 6.2) & (4.4, 12.2)& (0.5, 0.8) &\\
     \hhline{|=|=|=|=|=|=|}
     Onset to Death (\ref{fig:onset_to_death})&Inverse Gamma&19.0&14.6&$a=5.18$&0.006\\
     \hline
     C.I. (95\%) & & (16.6, 21.3) & (11.8, 25.0) &(3.5, 10.1) &\\
    \hhline{|=|=|=|=|=|=|}
     Hospitalization to Death (\ref{fig:hosp_to_death})&Gamma&16.6&15.8&$a=1.11$&0.008\\
     \hline
     C.I. (95\%) & &(13.9, 19.6) & (13.4, 21.7) & (0.6, 1.8)&\\
    \hhline{|=|=|=|=|=|=|}
     Hospitalization to Discharge (\ref{fig:hosp_to_discharge})&EMG& 17.6&12.5&$K=2.74$&0.004\\
     \hline
     C.I. (95\%) & &(16.9, 18.6)&(11.3, 13.5)& (2.3, 3.6)&\\
     \hhline{|=|=|=|=|=|=|}
     Test to Death (\ref{fig:test_to_death}) & Beta & 15.2 & 12.6 & $a=0.88$\newline $b=2.55$&$0.012$\\
     \hline
     C.I. (95\%) &  &$(12.2, 19.3)$& $(10.5, 15.4)$& $(0.7, 1.1)$\newline $(1.5,5.0)$ & \\
     \hline
    \end{tabular}
    \caption{Summary of best-fit distributions. }
    \label{table:dist_details}   
\end{table}

We note that some of the distributions are supported for ``negative" day delays as it was likely for a positive test to be performed before the onset of symptoms. 

From Fig. \ref{fig:distributions} and Table \ref{table:dist_details} we can see several significant trends in the course of the illness. Regarding hospitalizations, we see that regardless of the outcome of the illness, patients are likely to spend a long time in the hospital (likely between twenty days and a month), which suggests that the compilation effect, should the virus spread to a large part of the population, might have a detrimental effect to the healthcare system. Additionally, we see that patients are on average significantly later than their symptom onsets. Given that one is most contagious during the first few days of exhibiting symptoms \cite{cevik2020sars}, it is up to the patient to quarantine themselves before the positive result, and from a policy perspective, lowering that number (performing more tests sooner) may have a significant impact on the spread of the disease. Timing and availability of tests has become significantly better throughout the course of the pandemic, and it is likely for future extended data to reflect that trend.

\subsection{Imputation Results}
\begin{figure}[H]
	\centering
	\includegraphics[keepaspectratio,scale=0.6]{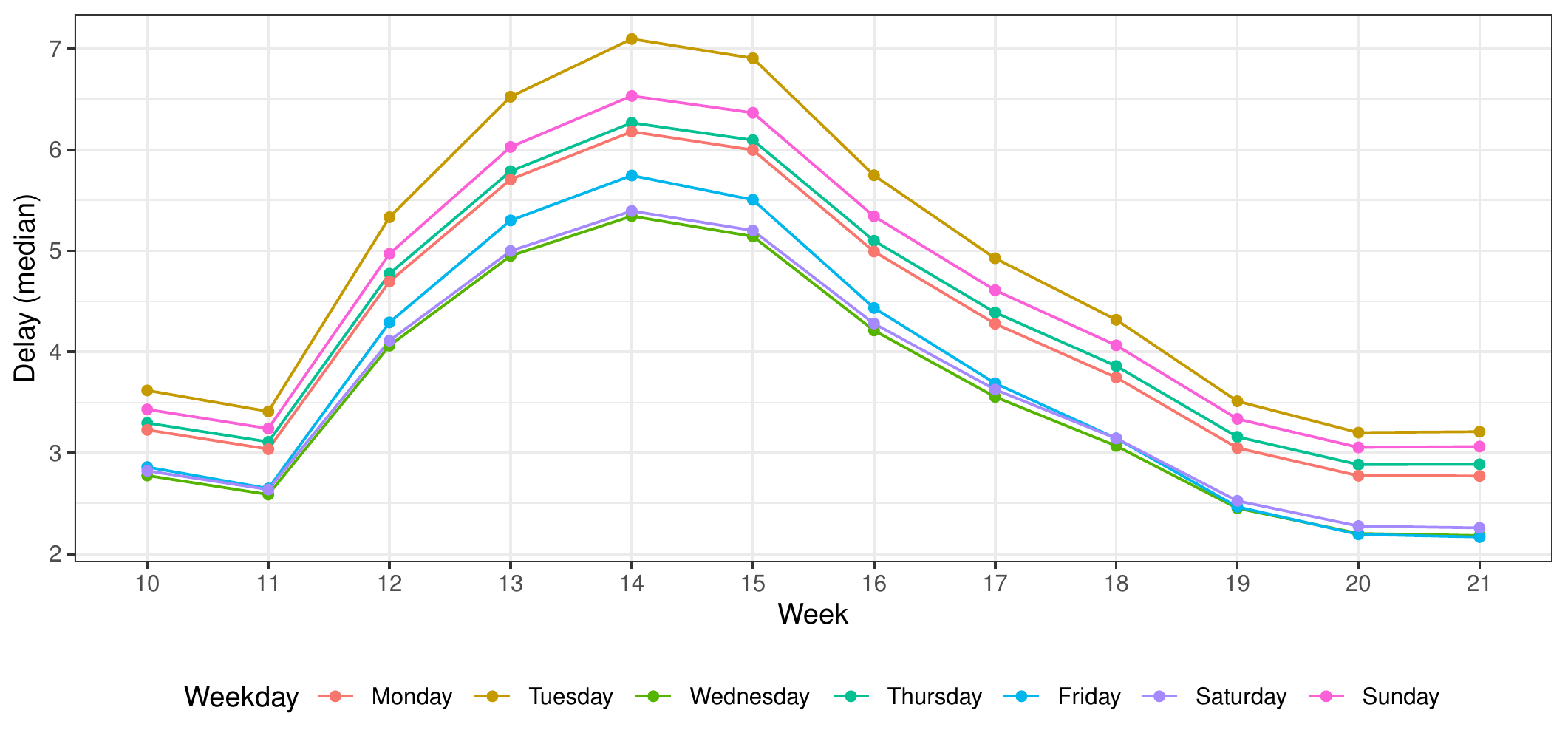}
	\caption{Results of the Weibull GAMLSS imputation model. Estimated median of the	delay time given case-specific covariates (reporting week, weekday of reporting).}\label{fig2}
\end{figure}

The association of the covariates with the median delay time of the fitted Weibull GAMLSS model is presented in Fig. \ref{fig2}. We find an increase of the median delay time from the 11th until the 15th week, followed by a decrease until the 20th week. We note that the estimated week effect is in agreement with the empirical median delay values. Moreover, we can see differences among the different days of the week (e.g. higher/lower median delays for cases reported on Tuesdays/Wednesdays). Details regarding the model diagnostics are presented in the supplementary material.

\begin{figure}[H]
	\centering
	\includegraphics[keepaspectratio,scale=0.85]{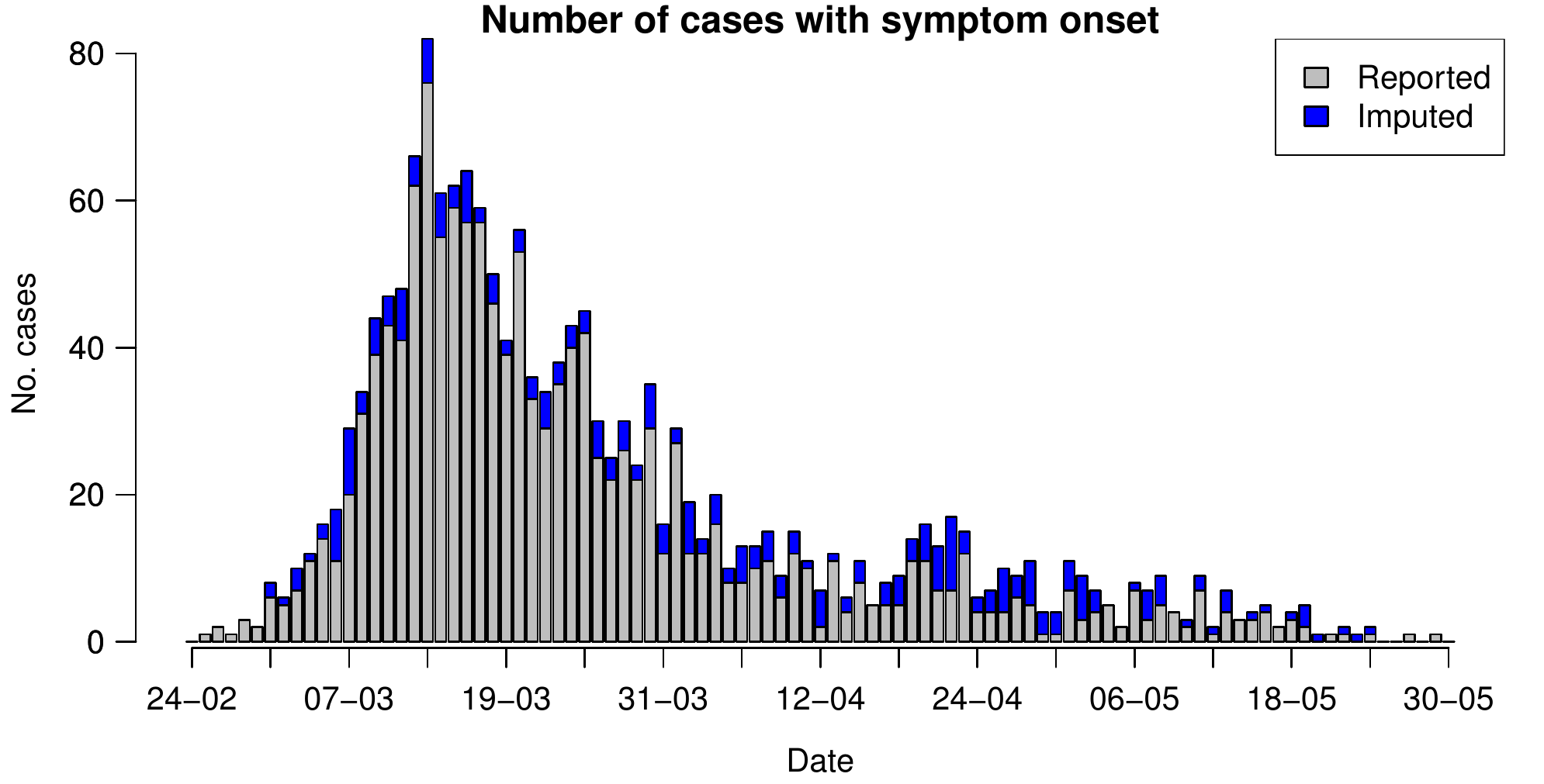}
	\caption{Imputation of the Attica CoViD--19 data, from 24/02/2020 until 30/05/2020. We present the observed (gray) and imputed (blue) numbers of symptom onset cases.}\label{fig3}
\end{figure}

We used the fitted Weibull GAMLSS model in order to impute the disease onsets, thus leaving the other observations unaffected. In Fig. \ref{fig3} we present the reconstructed epidemic curve depicting both the observed and imputed numbers of cases with symptom onset for each day between 24/02 and 30/05.

\subsection{Results for the effective reproduction number $R_e$}

\subsubsection{Estimation of $R_e$ by the Wallinga and Teunis method}

\begin{figure}[H]
	\centering
	\includegraphics[keepaspectratio,scale=0.65]{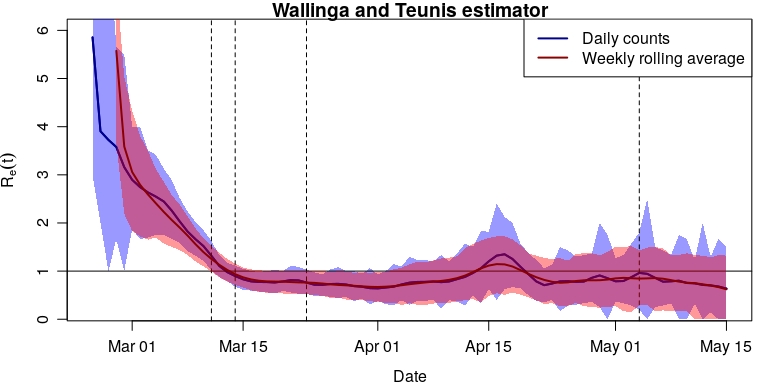}
	\caption{Wallinga and Teunis (time-dependent) \cite{wallinga2004different} estimator of $R_e$ for the daily data(blue)  and the weekly rolling average (red).} \label{fig:w&t}
\end{figure}

Our estimations for $R_e(t)$ by the Wallinga and Teunis (Time-Dependent) method are presented in Table \ref{tab1} and Fig. \ref{fig:w&t}, together with their 95\% confidence intervals. For the calculations we have used the R package ``R0'' \cite{obadia2012r0}. The method was applied not only to the daily data, but also to the associated time series obtained by taking a weekly rolling average. In Table \ref{tab1} we present the effective reproduction number estimates at 15-day intervals together with their $95\%$ confidence intervals (CIs), for both applications of the method.

\begin{table}
	\begin{center}
		\begin{tabular}{|c | c c c c c c|} 
			\hline
			Date $(t)$ & 01/03 & 15/03 & 01/04 & 15/04 & 01/05 & 15/05 \\ [0.5ex] 
			\hline
			\hline
			$\hat{R}_e(t)\,\textlatin{(daily)}$ & 2.90 & 0.83 & 0.64 & 1.2 & 0.79 & 0.64 \\ [0.5ex] 
			\hline 
			$\left(\hat{R}^{l}_e(t),\hat{R}^{u}_e(t)\right)$ & (1.88,4.00) & (0.63,1.05) & (0.38,0.90) & (0.64,1.82) & (0.36,1.27)  & (0.00, 1.51) \\ [0.5ex] 
			\hline
			\hline
			$\hat{R}_e(t)\,\textlatin{(weekly r.av.)}$ & 3.05 & 0.87 & 0.67 & 1.11 & 0.85 & 0.62   \\ [0.5ex] 
			\hline 
			$\left(\hat{R}^{l}_e(t),\hat{R}^{u}_e(t)\right)$ & (1.83,4.33) & (0.66,1.07) & (0.36,0.98) & (0.48,1.69) & (0.27,1.51) & (0.00,1.32) \\ [0.5ex] 
			\hline
		\end{tabular}
	\caption{Estimated $R_e(t)$  (effective reproduction number) with the associated $95\%$ confidence intervals using WT method, for the daily data and their weekly rolling average.} \label{tab1}
	\end{center}
\end{table}

In Fig. \ref{fig:w&t} we present the estimates for $R_e(t)$, superimposed with the $95\%$ CIs, obtained by $5,000$ simulation runs. We indicate with dashed lines the dates corresponding to the closure of schools and educational institutions (11/03), closure of restaurants, cafeterias and bars (13/03), as well as the beginning (23/03) and end (03/05) of the lockdown period.

The estimated $R_e$ fell below 1 for the first time at 13/03, and remained below 1 until 13/04, where we can see a rise above 1 from 14/04 to 19/04. From 20/04 and onwards until 15/05, the estimated $R_e$ did not exceed 1. The estimated $R_e(t)$ is in agreement between the daily data and the weekly rolling averages. Notably, the rolling averages result in a smoother $R_e$ curve, with lower uncertainty compared to the daily counts.

\subsubsection{Estimation of $R_{e}(t)$ by the Cori et al. method}

\begin{figure}[H]
	\centering
	\includegraphics[keepaspectratio,scale=0.7]{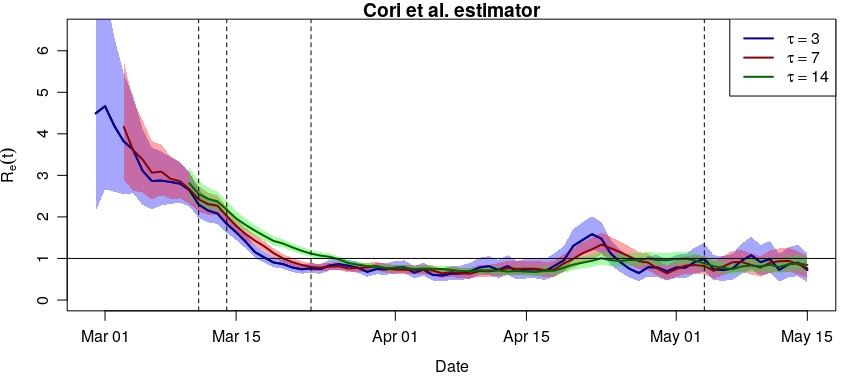}
	\caption{$R_{e}(t)$ \cite{cori2013new} (median) estimator from EpiEstim \cite{coriepiestim} package for different values of the smoothing window $\tau$, superimposed with the 95\% credible intervals.}\label{irn2}
\end{figure}

For the $R_{e}(t)$ estimations using the Cori et al. method, we have used the R package ``EpiEstim'' \cite{coriepiestim}, choosing 3 different widths of the smoothing window, namely $\tau \in \left\{3,7,14\right\}$ days. We note that, following Cori et al. \cite{cori2013new}, we report $R_{e,\tau}(t)$ for $t$ corresponding to the end of the window. The obtained estimates and their 95\% credible intervals are illustrated in Table \ref{tab2} and Fig. \ref{irn2}. 

\begin{table}[H]
	\begin{center}
		\begin{tabular}{|c || c c c c c c|} 
			\hline
			Date $(t)$ & 01/03 & 15/03 & 01/04 & 15/04 & 01/05 & 15/05 \\ [0.5ex] 
			\hline
			\hline 
			$\hat{R}_{e,\tau}(t),\, \tau=3$ & 4.67 & 1.63 & 0.78 & 0.73 & 0.78 & 0.72  \\ [0.5ex] 
			\hline 
			$\left(\hat{R}_{e,\tau}^{l}(t),\hat{R}_{e,\tau}^{u}(t)\right)$ & (2.67,7.47) & (1.44,1.83) & (0.64,0.94) & (0.52,1.00) & (0.54,1.09) & (0.43,1.12)  \\ [0.5ex] 
			\hline
			\hline
			$\hat{R}_{e,\tau}(t),\, \tau=7$ & - & 1.79 & 0.73 & 0.75 & 0.74 & 0.77  \\ [0.5ex] 
			\hline 
			$\left(\hat{R}_{e,\tau}^{l}(t),\hat{R}_{e,\tau}^{u}(t)\right)$ & - & (1.62,1.96) & (0.63,0.84) & (0.59,0.94) & (0.56,0.95) & (0.53,1.06)  \\ [0.5ex] 
			\hline
			\hline
			$\hat{R}_{e,\tau}(t),\, \tau=14$ & - & 1.96 & 0.77 & 0.68 & 0.9 & 0.84  \\ [0.5ex] 
			\hline	
			$\left(\hat{R}_{e,\tau}^{l}(t),\hat{R}_{e,\tau}^{u}(t)\right)$ & - & (1.80,2.13) & (0.71,0.84) & (0.59,0.78) & (0.83,1.15) & (0.67,1.04)  \\ [0.5ex] 
			\hline
	\end{tabular}
		\caption{Estimated $R_{e,\tau}(t)$ using the method of Cori et al.  \cite{cori2013new} for different sizes $\tau$ of the sliding window.}\label{tab2}
	\end{center}
\end{table}   

The increase of the window width leads to higher delays between the associated estimates and more smoothed curves. Due to the trade-off between noise and oversmoothing, a reasonable choice would be $\tau = 7$.

As we can see in Fig. \ref{irn2} initially there is a high $R_{e,\tau}$ (probably a result of overestimation \cite{cori2013new,gostic2020practical}), followed by a gradual decrease below 1 before the lockdown. In particular, the estimated $R_{e,\tau}$ fell for the first time below 1 at 21/03 ($\tau=7$). For $\tau=3$ and $\tau=14$, the associated dates of the first fall of $R_e$ below 1 are 19/03 and 26/03, respectively. During the lockdown, we observe an increase of $R_{e,\tau}$ above 1 that comes before a stabilization of $R_{e,\tau}$ below 1 from 27/04 until 15/05 (with $\tau=7$). The same pattern is observed with the estimated $R_e$ by the Wallinga-Teunis method, with the difference that -as expected- the $R_{e,\tau}$ estimates follow in time the WT estimates. Specifically, the periods during the lockdown in which the estimated effective reproduction numbers exceed 1 are 14/04-19/04 for WT and 21/04-26/04 for Cori et. al, (with $\tau=7$). 

The trade-off between high noise and oversmoothing is illustrated by the estimates obtained by $\tau=3$ and $\tau=14$. Specifically, for $\tau=3$ the $R_e$ curve is sensitive to small changes of the daily counts, leading to high fluctuations, especially at the beginning of May where e.g. $\hat{R}_{e,\tau=3} = 1.07$ at 09/05, with CI: (0.68,1.24). On the contrary, the $\hat{R}_{e,\tau=14}$ estimates remain below 1 from 27/03 until 15/05, with only exception a value of $\hat{R}_{e,\tau=14} = 1.005$ at 23/04, with CI: (0.86,1.17).

Two main observations are the decrease of $R_e(t)$ below $1$ before lockdown measures were imposed, as well as the aforementioned increase in the middle of April. The first observation has to be related with the decrease of the population mobility, while the second is probably related to a peak of the notified cases due to increase testing of asymptomatic cases in the second half of April (e.g. clusters at private healthcare facilities, testing of contacts).

\subsection{Sensitivity Analysis}
\begin{figure}[H]
	\centering
	\includegraphics[keepaspectratio,scale=0.8]{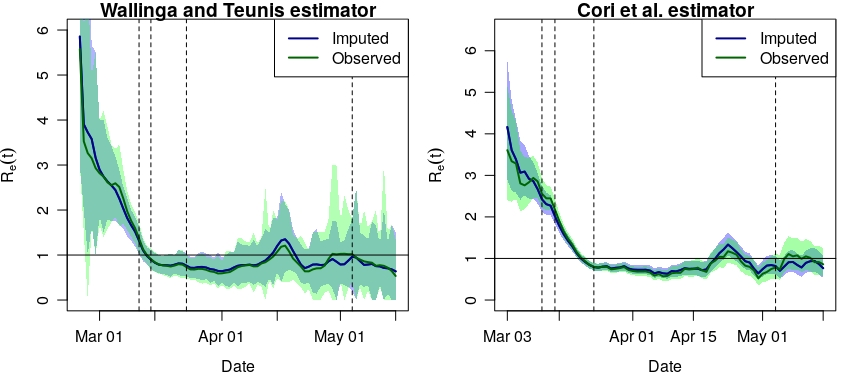}
	\caption{Sensitivity analysis: WT and Cori et al. estimations of $R_e$ with the associated 95\% uncertainty intervals, based on the imputed cases (blue), and non-imputed data (green).}\label{sa}
\end{figure}

In order to assess the impact of imputation, we have repeated the computations with WT and Cori et al. methods using as input data the available number of cases with known date of symptom onset, without performing imputation. The results of this analysis are shown in Fig. \ref{sa}, where we superimpose the WT and Cori et al. estimators (with a smoothing window width $\tau=7$). Indicative points together with their uncertainty intervals are reported in Table \ref{tab3}.

\begin{table}[H]
	\begin{center}
		\begin{tabular}{|c || c c c c c c|} 
			\hline
			Date $(t)$ & 01/03 & 15/03 & 01/04 & 15/04 & 01/05 & 15/05 \\ [0.5ex] 
			\hline
			\hline
			$\hat{R}_e(t),\,w_r=1$ & 2.82 & 0.84 & 0.59 & 1.07 & 1.02 & 0.53  \\ [0.5ex] 
			\hline 
			$\left(\hat{R}_e^{l}(t),\hat{R}_e^{u}(t)\right)$ & (1.88,4.00) & (0.63,1.05) & (0.38,0.93) & (0.64,1.82) & (0.36,1.27) & (0.00,1.51)  \\ [0.5ex] 
			\hline
			\hline
			$\hat{R}(t),\, w_s=7$ & - & 1.90 & 0.70 & 0.74 & 0.63 & 0.86   \\ [0.5ex] 
			\hline
			$\left(\hat{R}^{l}(t),\hat{R}^{u}(t)\right)$ & - & (1.72,2.10) & (0.60,0.81) & (0.56,0.96) & (0.43,0.89) & (0.56,1.24)  \\ [0.5ex] 
			\hline
		\end{tabular}
		\caption{Estimated $R_e(t)$ (effective reproduction number) and $R(t)$ (instantaneous reproduction number) for the observed (not imputed) data, with 95\% uncertainty intervals.}\label{tab3}
	\end{center}
\end{table}   

Regarding the non-imputed data WT estimations, we observe a fall of $R_e$ below 1 at 13/03, followed by an increase above 1 during the interval from 15/04 until 18/04, which is similar to the imputation-based results. However, we notice an additional rise of  $R_e$ between 29/04-04/05, followed by a steady decline. This is probably due to the small number of available cases, as indicated by the high level of uncertainty (e.g. at 29/04 the confidence interval ranges from 0.17 to 1.5). A similar behavior is also observed at the non-imputed estimates with the Cori et al. method. While the estimated curve of $R_e$ is in agreement with the imputed-based results, we find an additional increase above 1 during 07/05-12/05. In summary, the imputation process did not make qualitatively significant changes on the estimates, while it led to more consistent results with lower levels of uncertainty.\par
 
While it is safer to say that the measures had a significant role in consistently keeping $R(t)$ below $1$ in the months following their implementation, data-driven estimators may have difficulty capturing their effect (in conjunction with estimating the general $R_e(t)$) during the beginning stages of the spread. One way to possibly retrospectively correct this is to use data from the second and third waves of the pandemic in several countries (Greece included) during which testing was more widespread and accurate; making reversed-time projections might give us different estimates for the true number of cases in February or March.\par

\begin{figure}[H]
	\centering
	\includegraphics[keepaspectratio,scale=0.8]{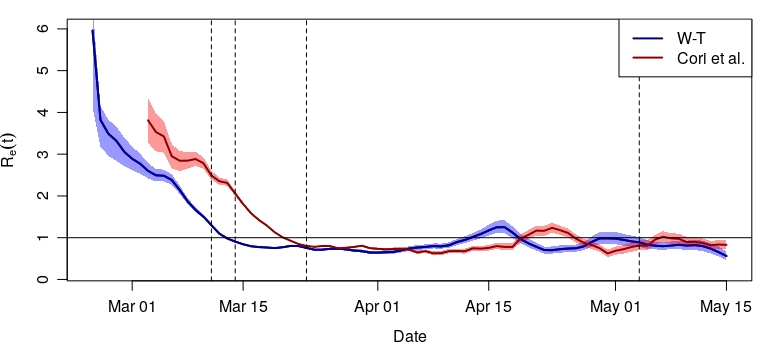}
	\caption{Sensitivity analysis: Medians of and WT and Cori et al. $R_e(t)$ estimations, with the associated 95\% uncertainty intervals, based on 1,000 imputed data sets.}\label{uq}
\end{figure}

We are also interested in examining the effect of the uncertainty stemming from the imputation process on the obtained estimators. To this end, we generate 1,000 imputed data sets (sampling for each data set the delay times from the fitted Weibull distribution) and calculate the $R_e(t)$ estimations for both WT and Cori et al. ($\tau=7$) methods. From the obtained 1,000 estimations for each day, we calculate the medians and present them in Fig. \ref{uq}, together with the uncertainty intervals constructed by the associated 2.5\% and 97.5\% quantiles (for both methods). \par

\begin{figure}[H]
	\centering
	\includegraphics[keepaspectratio,scale=0.8]{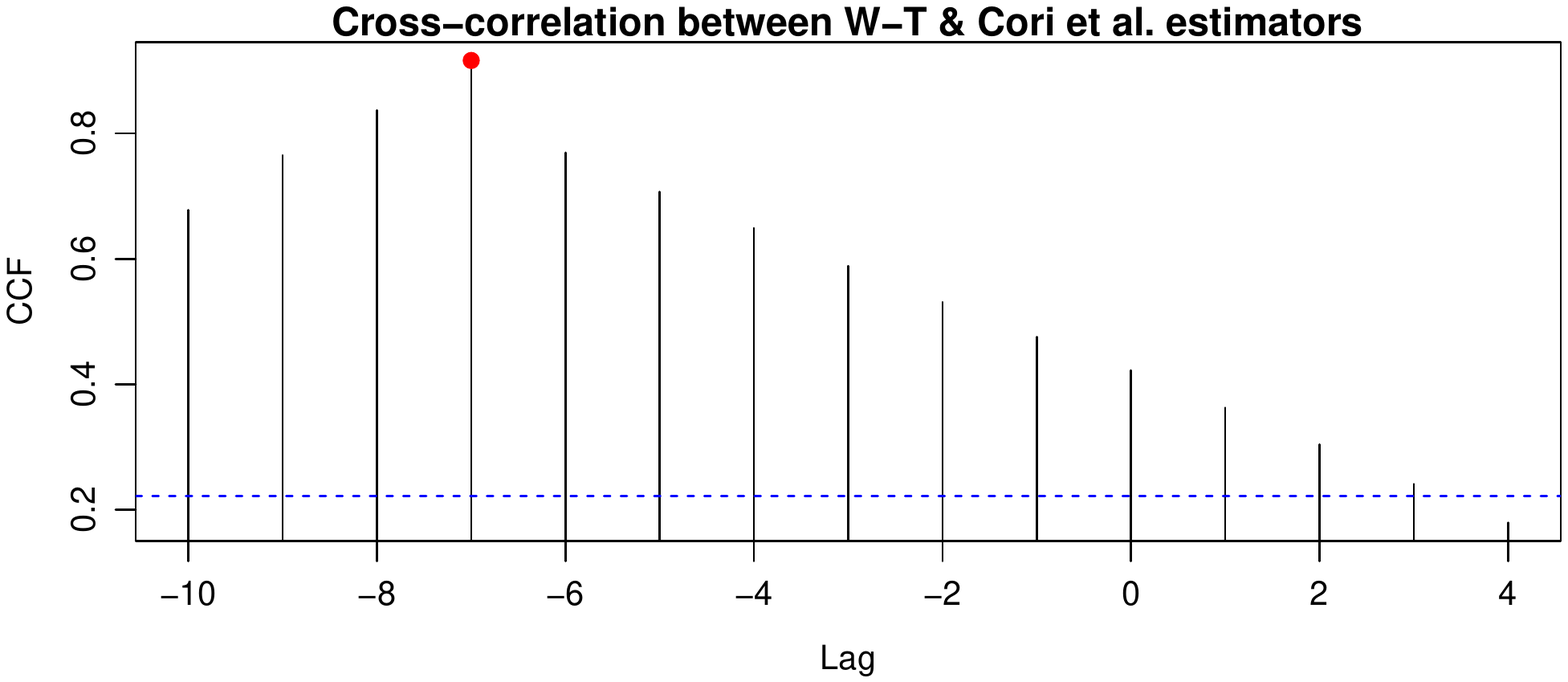}
	\caption{Cross-correlation function between WT and Cori et al. estimations of $R_e(t)$. We indicate with a red circle the value of maximum correlation (0.92) at lag -7.}\label{ccf}
\end{figure}

As anticipated, the Cori et al. estimates follow in time the WT ones. In order to further examine this relation, we present in Fig. \ref{ccf} the cross-correlation function \cite{shumway2017time} between WT and Cori et al. estimations of $R_e(t)$ (displayed in Fig. \ref{uq}). The peak at lag equal to -7 with correlation 0.92 means that there is a statistically significant strong positive correlation between the two time series, with the Cori et al. estimates following the WT ones, which are 7 days ahead in time. One main reason for that, is that the WT estimations are defined forward in time, contrary to the Cori et al. $R_e(t)$ results, which use observations up to time $t$.

\section{Discussion and Conclusions}

Since early 2020, the world has been facing a pandemic with unprecedented health, economic and societal consequences globally. In order to limit person-to-person transmission and contain the pandemic, strict social distancing measures, including nationwide lockdowns, were implemented is most countries. \par

To the best of our knowledge, this is the first study to estimate the $R_e$ factor during the first COVID-19 epidemic wave in Greece. This knowledge is of utmost importance in order to guide future public health interventions, both at the national, but also at the regional level. For this reason, we performed an analysis using data of notified cases in the Attica region, where almost half of the country's population resides. It is important to note that Greece is characterized by high geographic heterogeneity regarding the SARS-CoV-2 spreading, especially in the early period. Thus, an analysis using data from the whole country may yield misleading or inaccurate results; many models are built around the assumption of a well-mixed population. We expect that the densely populated and relatively cohesive and homogeneous character of the population of the metropolitan region of Athens and its suburbs, compared to the rest of the country, is better suited for this type of analysis. Our analysis is based on the imputed number of cases with respect to dates of symptom onset and not on the number of notified cases.\par

In this work, we implemented two of the most widely used methods (that of Wallinga and Teunis and Cori et al. \cite{cori2013new}) for the estimation of the effective reproduction number for the first wave of the pandemic in the greater metropolitan area of Athens, Greece. Similar studies regarding the estimation of the effective reproduction number have been performed for the cases of Bavaria \cite{gunther2020nowcasting}, Spain \cite{santamaria2020covid}, Latin American countries \cite{ochoa2020effective}, the Philippines \cite{haw2020epidemiological}, and South Korea \cite{ryu2020effect}, among others.  \par

As a first step we imputed the dates of symptoms onset based on the dates of notified cases. For that purpose, we used the GAMLSS model to fit a Weibull distribution to approximate the delay between the time of onset of symptoms and the time of reporting. The imputation provided evidence of significant effects of the reporting week and weekday. The Wallinga \& Teunis and Cori et al. gave qualitatively similar results. In the first period from February 26 to early March, both methods estimated relatively high $R_t$ and high uncertainty that should be attributed to the very low numbers of notified cases. Thus, on March 1, the $R_e$ was estimated to be around 3 (95\% CI: ~1.80,~4.3). Our analysis suggests that $R_e$ dropped below 1 around March 15. In particular, on March 15 $R_e$ was around 0.85 (95\% CI: ~0.65,1.05). This is also in line with the result reported in Lytras et al.\cite{lytras2020improved} for the while country that. In particular, they also find that the $R_e$ dropped below 1 one week before the implementation of the full lockdown. However, both methods that we used for our analysis showed an increase of $R_t$ during the lockdown, which should be attributed to an increase of notified cases in specific clusters. On April 15, both methods resulted in an expected value of $R_e$ above 1, but the results are characterized by relatively high uncertainty. More specifically, the Wallinga \& Teunis method estimated $R_e=1.2$ (95\% CI: ~0.64,~1.82)  and the Cori et al. method estimated $R_e=1.2$ (95\% CI: ~0.48,~1.69). After that date, both methods in agreement estimated the $R_e$ to be below 1, but with the upper bounds still above 1. The above findings were quantitatively similar also when we considered just the notified cases, i.e. without imputation. \par
We should note that the above analysis should be viewed critically as one has to consider the validity of the application of both methods for the calculation of $R_e$. First, both methods do not take into account the imported cases (although the methods can be modified to also take into account imported cases). Thus, such an assumption may be have been violated for the first period before the lockdown. However, as (a) the period of February is not a touristic one for Greece, (b) the pandemic has already limited traveling by early February, we expect that the violation of this assumption is not critical for the results. In fact, based on the statistics released by Eurostat \cite{Eurostat}, the drop of the number of nights spent in tourist accommodation establishments in the period January-February 2020 (compared with the same period of 2019) was of the order of 43\% in Greece, which was the biggest drop among the countries of EU27. Another important assumption of both methods is that the infection  network  can be constructed based only on the notified cases. However, in general, due to under-reporting,  this assumption can hardly be met. On the other hand, as in Greece and in the greater metropolitan area of Athens in particular, the number of cases was at very low levels during the first wave of the pandemic, the above assumption could be considered to be partly valid. Finally, we note that the results of such analyses for the estimation of the $R_e$ should be taken with caution for policy making as the assumptions underlying their implementation can be easily violated, mostly due to the under-reporting.

\section{Acknowledgments}

We acknowledge the National Public Health Organization (EODY) for providing us with the detailed epidemiological data for Attica, Greece.

\section*{Supplementary information}

We provide here additional information related to the fitting of the GAMLSS model for the imputation step. Regarding the data used to fit the model, we use the notified cases with calendar week number ranging from 10 until 21, with non-negative delay time $t_d \leq 20$. For the cases where $t_d=0$ days, we transform them as $t_d=0.5$ days. This is not only due to the necessary log-transformation, but also because practically the delay time cannot be exactly 0. 

Our code was based on the code provided by the github public repository\footnote{\url{https://github.com/FelixGuenther/nc_covid19_bavaria}} for the manuscript \cite{gunther2020nowcasting}. For the sake of completeness, we provide below the output produced by R. The associated p-values for the calendar week number (rep\_week\_local) and the specific weekdays (rep\_date\_local\_weekday) are presented in the last column.

\begin{verbatim}
------------------------------------------------------------------
Mu link function:  log
Mu Coefficients:
                                Estimate Std. Error t value Pr(>|t|)    
(Intercept)                     -2.47004    0.36592  -6.750 2.24e-11 ***
pb(rep_week_local)               0.02407    0.02483   0.969   0.3326    
rep_date_local_weekdayMonday    -0.38052    0.22947  -1.658   0.0975 .  
rep_date_local_weekdaySaturday  -0.11905    0.24382  -0.488   0.6254    
rep_date_local_weekdaySunday    -0.52781    0.30600  -1.725   0.0848 .  
rep_date_local_weekdayThursday  -0.45039    0.22921  -1.965   0.0496 *  
rep_date_local_weekdayTuesday   -0.52249    0.24251  -2.154   0.0314 *  
rep_date_local_weekdayWednesday -0.06606    0.22160  -0.298   0.7657    
---
Signif. codes:  0 ‘***’ 0.001 ‘**’ 0.01 ‘*’ 0.05 ‘.’ 0.1 ‘ ’ 1

------------------------------------------------------------------
Sigma link function:  log
Sigma Coefficients:
                                 Estimate Std. Error t value Pr(>|t|)    
(Intercept)                      0.639010   0.131340   4.865 1.29e-06 ***
pb(rep_week_local)              -0.029876   0.008993  -3.322 0.000919 ***
rep_date_local_weekdayMonday     0.120129   0.074188   1.619 0.105640    
rep_date_local_weekdaySaturday   0.089936   0.083196   1.081 0.279895    
rep_date_local_weekdaySunday     0.145929   0.096941   1.505 0.132486    
rep_date_local_weekdayThursday   0.139354   0.073916   1.885 0.059617 .  
rep_date_local_weekdayTuesday    0.100802   0.077483   1.301 0.193510    
rep_date_local_weekdayWednesday  0.072260   0.075882   0.952 0.341142    
---
Signif. codes:  0 ‘***’ 0.001 ‘**’ 0.01 ‘*’ 0.05 ‘.’ 0.1 ‘ ’ 1

------------------------------------------------------------------
No. of observations in the fit:  1288 
Degrees of Freedom for the fit:  22.59269
      Residual Deg. of Freedom:  1265.407 
                      at cycle:  4 
 
Global Deviance:     6611.329 
            AIC:     6656.514 
            SBC:     6773.112 
******************************************************************
\end{verbatim}

\begin{figure}[H]
	\centering
	\includegraphics[keepaspectratio,scale=0.8]{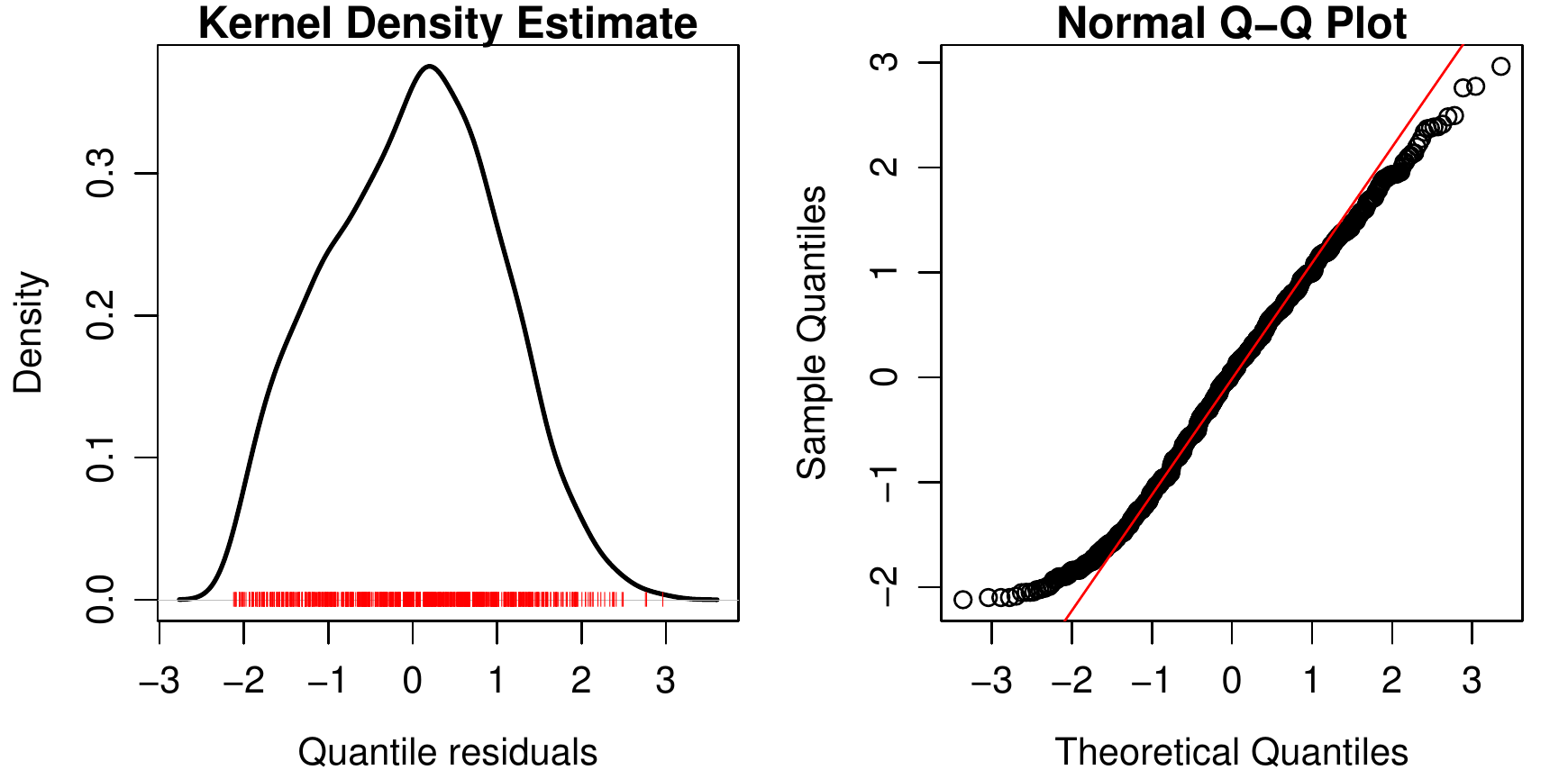}
	\caption{KDE (left) and qq-plot (right) of the (normalized quantile) residuals of the fitted GAMLSS model.}\label{resdiag}
\end{figure}

In order to further examine the model fit, we present in Fig. \ref{resdiag} the  kernel density estimate (KDE) and the qq-plot of the (normalized randomized) quantile residuals.

We note that the the quantile residuals of this model behave well, e.g. their
mean is nearly zero (-0.002), their variance nearly one (1.00), their coefficient of skewness near zero (0.0177) and their coefficient of kurtosis is near 3 (2.44). Taking also into account the residuals KDE and qq-plot presented in Fig. \ref{resdiag} we conclude that we have an approximately normal distribution, indicative of an adequate model.

\begin{figure}[H]
	\centering
	\includegraphics[keepaspectratio,scale=0.6]{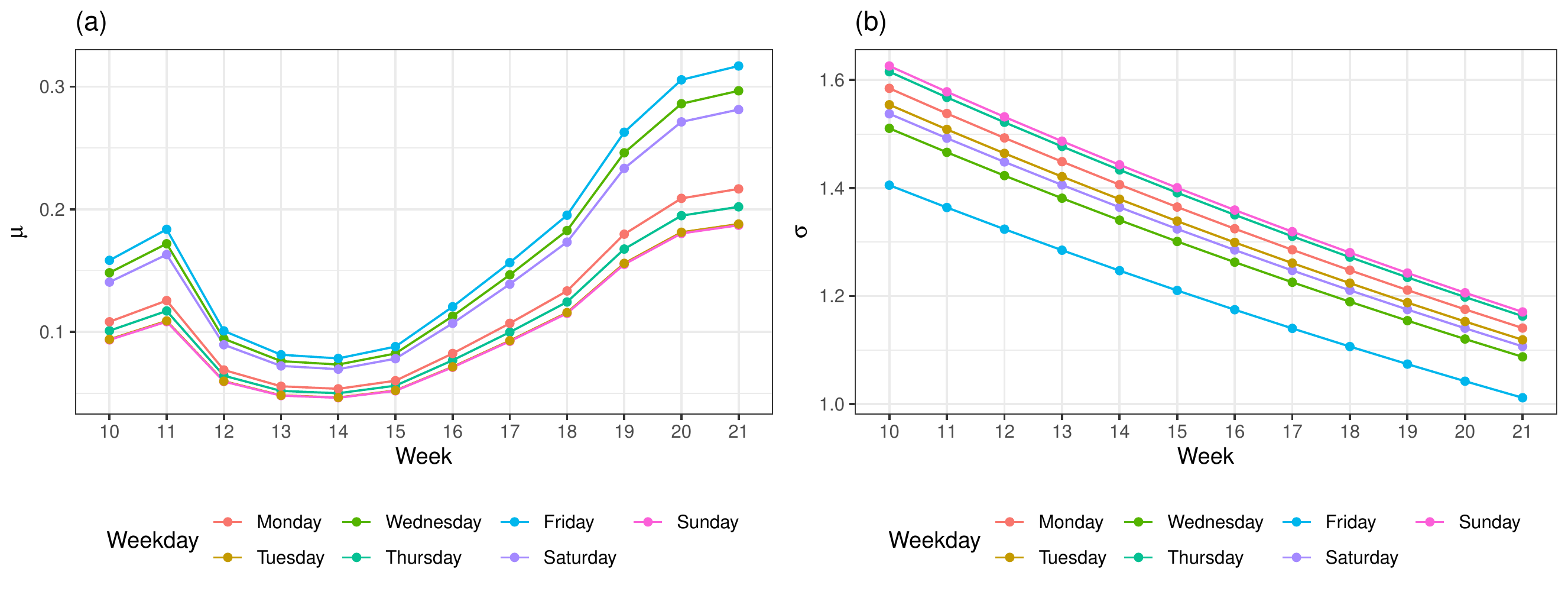}
	\caption{Fitted parameters $\mu$ and $\sigma$ of the delay time Weibull distribution with respect to the weekday and the week number.}\label{mu_sigma}
\end{figure}

Finally, in Fig. \ref{mu_sigma} we present the obtained parameters $\mu$ and $\sigma$ of the delay time Weibull distribution with respect to the weekday and the week number.

\bibliographystyle{unsrt}
\bibliography{covid_report.bib}    

\end{document}